\documentclass[a4paper,11pt]{article}
\pdfoutput=1 % if your are submitting a pdflatex (i.e. if you have
\usepackage{jinstpub} % for details on the use of the package, please
\usepackage{epstopdf}
\usepackage{gensymb}

\title{Performance verification of the CMS Phase-1 Upgrade Pixel detector}

\author[a,1,2]{Viktor Veszpremi\note{for the CMS Tracker group}\note{Supported by the Janos Bolyai Fellowship of the Hungarian Academy of Sciences}}

\affiliation[a]{Winger Research Centre for Physics,\\Budapest, Hungary}

\emailAdd{veszpremi.viktor@wigner.mta.hu}

\abstract{The CMS tracker consists of two tracking systems utilizing semiconductor technology: the inner pixel and the outer strip detectors. The tracker detectors occupy the volume around the beam interaction region between 3\,cm and 110\,cm in radius and up to 280\,cm along the beam axis. The pixel detector consists of 124 million pixels, corresponding to about 2\,m\,$^{2}$ total area. It plays a vital role in the seeding of the track reconstruction algorithms and in the reconstruction of primary interactions and secondary decay vertices. It is surrounded by the strip tracker with 10 million read-out channels, corresponding to 200\,m\,$^{2}$ total area. The tracker is operated in a high-occupancy and high-radiation environment established by particle collisions in the LHC. The performance of the silicon strip detector continues to be of high quality. The pixel detector that has been used in Run 1 and in the first half of Run 2 was, however, replaced with the so-called Phase-1 Upgrade detector. The new system 
is 
better suited to match the increased instantaneous luminosity the LHC would reach before 2023. It was built to operate at an instantaneous luminosity of around 2$\times$10$^{34}$\,cm$^{-2}$s$^{-1}$. The detector's new layout has an additional inner layer with respect to the previous one; it allows for more efficient tracking with smaller fake rate at higher event pile-up. The paper focuses on the first results obtained during the commissioning of the new detector. It also includes challenges faced during the first data taking to reach the optimal measurement efficiency. Details will be given on the performance at high occupancy with respect to observables such as data-rate, hit reconstruction efficiency, and resolution.}


\collaboration[c]{on behalf of the CMS Collaboration}

\proceeding{11$^{\text{th}}$ International Conference of Position Sensitive Detectors\\
  3-8 September 2017\\
  Open University, Milton Keynes, UK}

\begin{document}
\maketitle
\flushbottom

\section{Introduction}
\label{sec:intro}

The tracker detectors~\cite{TrackerTDR,TrackerTDRadd} of the Compact Muon Solenoid (CMS) experiment~\cite{CMSExperiment} are responsible for measuring trajectories of charged particles that are created in particle collisions inside the Large Hadron Collider (LHC). Charged particles are detected by multiple tracking layers causing the least amount of interaction achievable between the particles and the sensor modules. In the CMS tracker, sensor modules are arranged on surfaces of cylinders which are rotationally symmetric to the LHC beam line, the $z$ axis of the CMS reference frame. The lateral surfaces of the cylinders are called the barrel, while the bases are called the forward or endcap detectors. The pixel detector occupies the inner part of the tracker volume up to a radius of 16\,cm, situated directly around the LHC beam-pipe starting at 2.9\,cm. The barrel (BPix) is composed of four layers. The forward (FPix) is made of three disks on each side. The strip detector fully  surrounds the pixel detector. 

The main track reconstruction algorithms~\cite{Tracking} follow an inside-out logic. Trajectory seeds are formed from hit-multiplets in the pixel detector with configurations that are compatible with particles originating from the beam interaction region. Trajectories are built using measurements in the strip detector following an implementation of the Kalman filter method. The removal of the hits which are processed by the pixel-seeded steps give room for further algorithms. Algorithms exist for reconstructing muon tracks seeded outside-in by the CMS Muon system, electrons seeded from the calorimeters towards the pixel detector, and particles from secondary interactions originating from outside the pixel detector. The primary vertices in the beam interaction regions and secondary vertices from decays of sufficiently long lived particles are reconstructed as the intersections of multiple tracks.

Being closest to the beam-crossing region, the pixel detector, of all CMS subsystems, operates in the environment with the highest particle density. The detector is expected to return measurement points for all particles that are created in the beam-beam interactions at 40\,MHz rate upon receiving read-out trigger signals with typical frequencies up to 100\,kHz. The LHC is expected to deliver proton-proton collisions at 2.1$\times$10$^{34}$\,cm$^{-2}$s$^{-1}$ instantaneous luminosity before 2019, which corresponds to about 60 proton-proton (``pile-up'') interactions in each beam-crossing, and to reach 2.5$\times$10$^{34}$\,cm$^{-2}$s$^{-1}$ until 2023. The single hit efficiency in the innermost layer of the pixel detector was measured as 94.5\% at the peak luminosity of 1.45$\times$10$^{34}$\,cm$^{-2}$s$^{-1}$ in 2016, at a particle hit rate already 40\% over the design goal. The higher rate, along with the incurred radiation damage, made the upgrade of the pixel detector necessary. The replacement of the 
old detector with the so-called Phase-1 Upgrade detector~\cite{Phase1TDR} was carried out at the beginning of 2017. Technical details on the design and technological choices are reviewed elsewhere~\cite{Phase1DetectorProceedings}, merely a summary of the upgrade goals are given in the rest of this section. The commissioning and the preliminary steps of the calibration concluded at the end of Summer, 2017. The performance verification of the upgraded detector is presented in the following sections.

The higher number of pile-up interactions requires a better vertex separation. The vertex resolution is determined by the number of tracks used in the vertex fit and the impact parameter resolution of those tracks. The former parameter may be enhanced by reconstructing tracks at higher pseudo-rapidities and with lower transverse momentum. In the Phase-1 pixel detector, this is achieved by adding an extra layer in the barrel and an extra disk in the forward regions. The radii of the four barrel layers are 2.9\,cm, 6.8\,cm, 10.9\,cm, and 16.0\,cm. The three forward disks are positioned along $z$ at 3.2\,cm, 3.9\,cm, and 4.8\,cm. Each disk is composed of two rings of modules with average radii of 12.8\,cm and 7.8\,cm. Seeding from hit quadruplets is possible up to $|\eta|$\,<\,2.5 in pseudo-rapidity. The impact parameters of the tracks are improved by moving the innermost layer (Layer~1) closer to the beam and the outermost layer (Layer 4) closer to the first layer of the strip detector. It is important to note 
that the higher number of layers leads to doubling the number of modules, but it comes with a slight decrease in the overall material budget. 

\section{Read-out electronics and DAQ}

The read-out chips (ROC) of the pixel modules are supplied with digital and analog low voltages via the service cylinders located at the two ends of the detector cylinders. High voltage for the sensor bias, I$^{2}$C programming and command transmission, as well as the data read-out are also facilitated through the service cylinders. A pixel module is composed of 16 ROCs arranged in a 2-by-8 array of size 1.6\,cm by 6.3\,cm. The new detector is built on the same n$^{+}$-in-n sensor design as the old one. The pixel size is 100\,$\mu$m$\times$150\,$\mu$m, the thickness of the sensitive volume is 285\,$\mu$m. 

\begin{figure}[htbp]
\centering
\qquad
\includegraphics[width=.7\textwidth,origin=c]{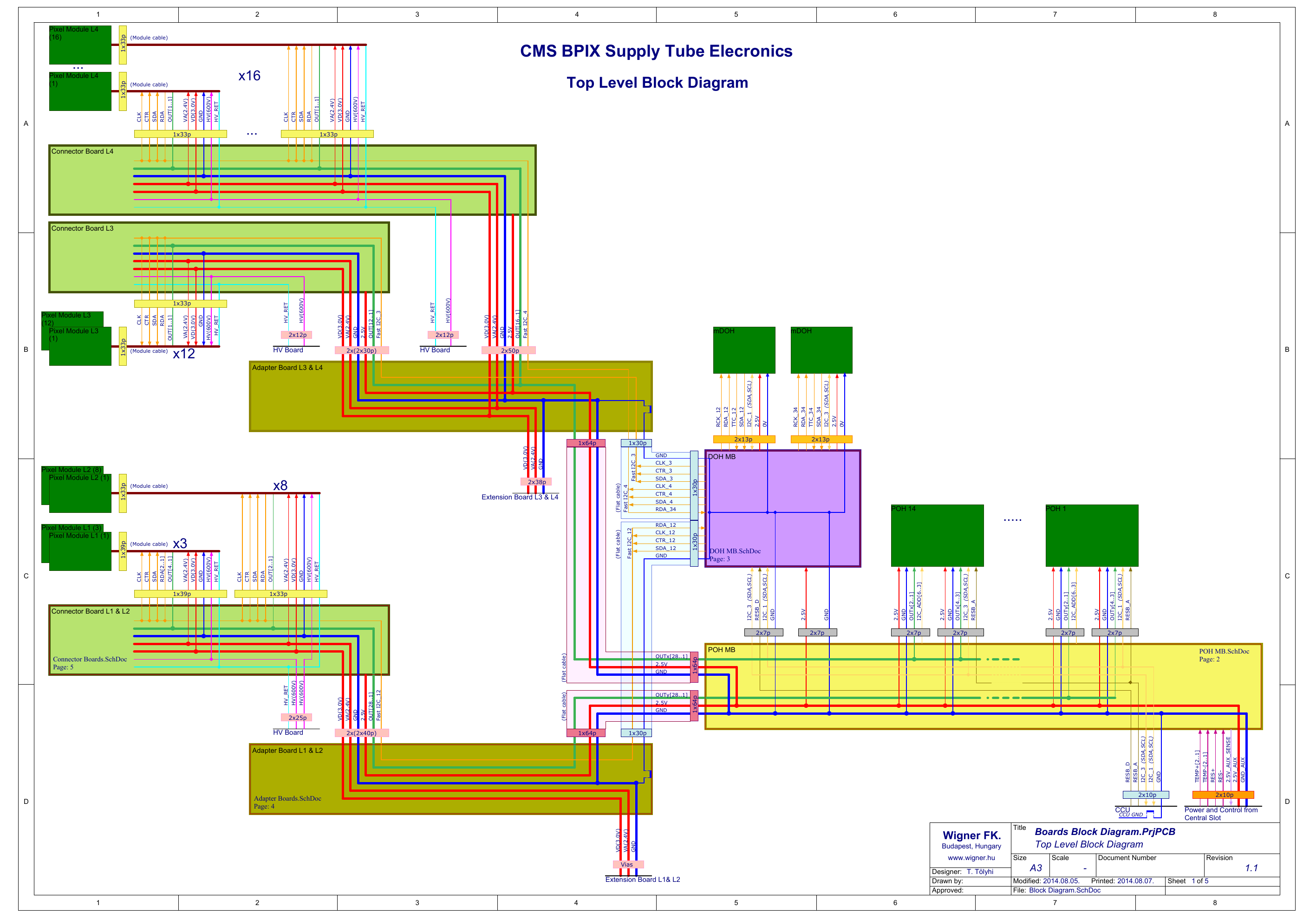}
\caption{\label{fig:LayersDisksServices} %Left: $r-z$ view of the four barrel layers, three forward disks, and the service cylinders. The services have been moved from the blue region to the green one in order to reduce the material budget. The right figure shows 
The printed circuit boards and their connections in the barrel supply tube. The sensor modules are represented by the dark green boxes on the left, the opto-hybrids (DOH and POH) by the dark green boxes on the right. }
\end{figure}

The detector modules are grouped into 16 sectors corresponding to roughly equal size wedges in the transverse plane. Each sector belongs to a set of printed circuit boards arranged in a slot of the service cylinders. Most of the electronics for the forward disk sectors are implemented in single cards of the same kind, called the Port-card. The structure of the slots in the barrel service cylinders are somewhat more complex, as shown in Figure~\ref{fig:LayersDisksServices}. Modules are connected to the Connector Boards (CB) through twisted pair cables. High-voltage is distributed through the CB. There are separate boards for Layer 3 and 4, and a combined one for Layer~1 and 2. Digital signals to and from the CB and the low-voltage supplies~\cite{Phase1DCDCConverters} are combined into two Adapter Boards (AB), one for Layer 3 and 4 and one for Layer~1 and 2. The two AB receive I$^{2}$C programming signals from a single board (DOH MB) that hosts digital opto-hybrid modules (DOH) facilitating optical 
communication with the Front-End Controller (FEC) cards at the back-end. The AB also transmit data to a board (POH MB) that carries opto-hybrids (POH)~\cite{Phase1POH} for optical read-out by the Frond-End Driver cards (FEDs) at the back-end. Both FED and FEC are FPGA boards in $\mu$TCA standard format.

FEDs receive data input in two connectors with 12 optical receivers on each connector. The data in a single fiber is transmitted via a 400\,MHz digital communication protocol, in which data from two 160\,MHz channels are overlaid. A single 160\,MHz channel transmits hits from two ROCs in a module in Layer~1, four ROCs in Layer 2, and eight ROCs in Layer 3 and 4, as well as in the forward disks. Since a module is composed of 16 ROCs, a Layer~1 module is read-out in eight channels, via four optical fibers. FEDs have a single 10\,Gbps output link.

\begin{figure}[htbp]
\centering
\includegraphics[width=.35\textwidth,origin=c]{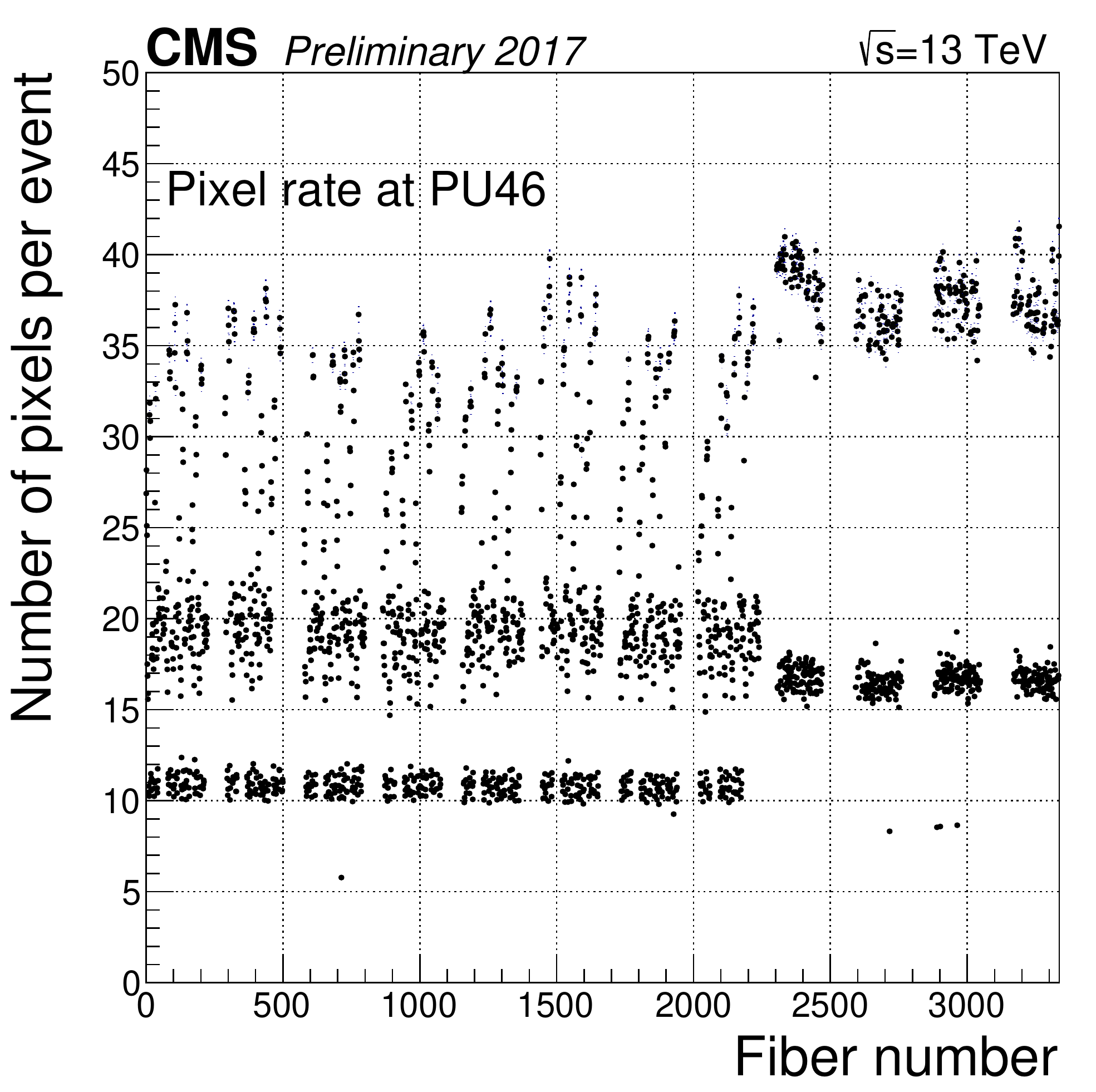}
\qquad
\includegraphics[width=.35\textwidth,origin=c]{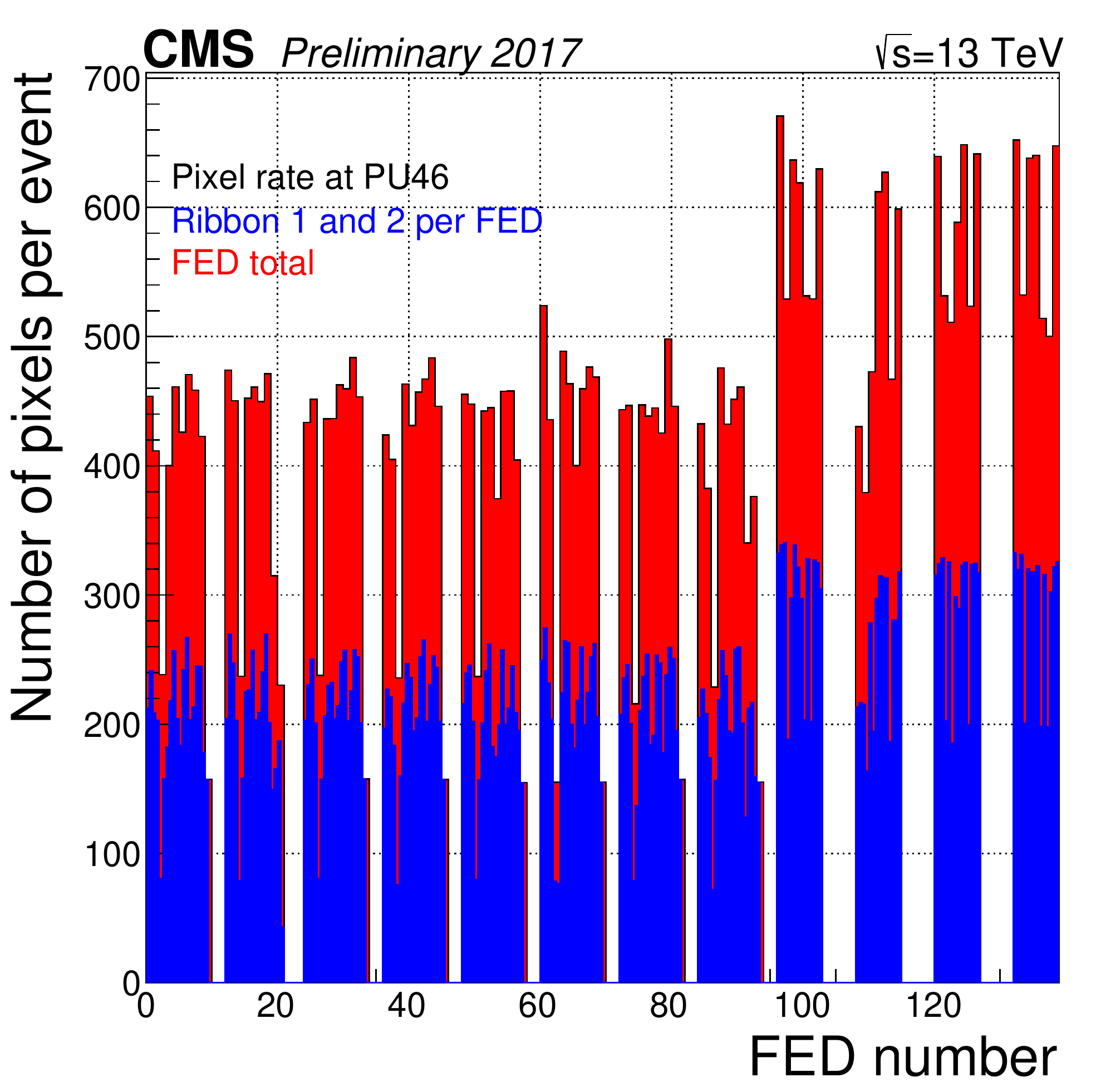}
\caption{\label{fig:PixelRates} Left: number of pixels per event in single fibers in the barrel and forward modules. Right: equalized total rate per FED (red) overlaid with rate in each of the two FED connectors (blue)~\cite{CMSPixelOfflinePublicPage}. }
\end{figure}

In Figure~\ref{fig:PixelRates}, the plot on the left shows the average number of pixels per event transmitted in every fiber to every FED in the pixel system when the average pile-up was 46. There are more FED identification numbers pre-allocated for the detector than the actual number of FEDs that are populated at the moment; therefore, there are fiber numbers with no measured hit rate. A clear separation in the pattern along the horizontal axis is visible between the BPix fibers with lower and FPix fibers with higher serial number. The two groups of measurement points along the vertical axis for the FPix fibers represent modules positioned at the two different radii in every disk, called the inner and outer rings. In BPix, fibers transmitting Layer 4 data see the lowest rate. Layer 2 and 3 fibers have similar rates, about twice of Layer 4. Layer~1 fibers transmit up to four times more hits than those for Layer 4. While a per disk read-out in FPix sectors 
balances out the rates between inner and outer rings, it is easy to see that a per layer parallel read-out in the BPix sectors would lead to a very uneven distribution of the load on the FEDs. The concentration of all the read-out channels on a single POH MB was exploited by implementing a load-balancing routing of the POH modules in such a way that the total rates on each FED connector are equalized. This is seen in Figure~\ref{fig:PixelRates} on the right. The red bars show the average number of pixels per event received in all FEDs, while the overlaid blue bars show the two connectors on the FED individually. The routing of the POH MB was established based on early Monte Carlo simulations of the Phase-1 detector. The FED with the highest load is found in the FPix.

The theoretical limit on the data-rate is considered to be due to the saturation of the FED input links. The present rate limit estimated from data is reached around 2.1$\times$10$^{34}$\,cm$^{-2}$s$^{-1}$ instantaneous luminosity determined by the current FED firmware. This meets the requirement until the next shut-down period of the LHC.

\section{Detector timing}
\label{sec:timing}

Sensor modules have been bump-bonded with two types of read-out chips. A modified version of the PSI46 chip called the PSI46dig, and an upgraded chip that was designed to operate at 600\,MHz/cm$^{2}$ particle rate, called PROC600~\cite{PROC600Testbeam}. In both ROCs, pixel hits are registered and buffered asynchronously along with their time-stamps awaiting for a triggered read-out. When the trigger command arrives, hits corresponding to a preset latency, roughly comparable to the latency of the trigger decision in CMS, are retrieved from the buffers. The unit of the time-stamp counter is the LHC clock period, approximately 25\,ns. The assignment of the time-stamp for a hit depends on the relative phase between the internal clock of the ROC and when the collision in the bunch crossing occurs. A phase-delay of the internal clock is ideally adjusted by maximizing the number of pixel hits generated by all particles in a triggered collision while also maximizing the sensor efficiency.

Due to the geometry of the barrel, and that it is built using two different kinds of ROC, the internal synchronization of the BPix modules required a certain amount of care. The clock and the trigger signals are distributed to the pixel modules from the DOH MB. There are two distribution lines for each slot: one grouping modules in Layer~1 and 2, and one for modules in Layer 3 and 4. The clock phase for a group is programmable by a delay chip on the DOH MB. The internal clock of modules on the same clock-distribution line cannot be delayed in any programmable way, it is determined by the signal delays in the hardware. The signal propagation time for each module was computed individually and adjusted such that modules receive their input clock with a time-difference corresponding to the average time of flight of particles arriving from the beam interaction region. The signal propagation time is determined by the module cable length, the position of the module connector on the CB, and the routing distance 
between the module connector and the DOH MB. The average expected time of flight was computed from early Monte Carlo simulations.

\begin{figure}[htbp]
\centering
\includegraphics[width=.35\textwidth,origin=c]{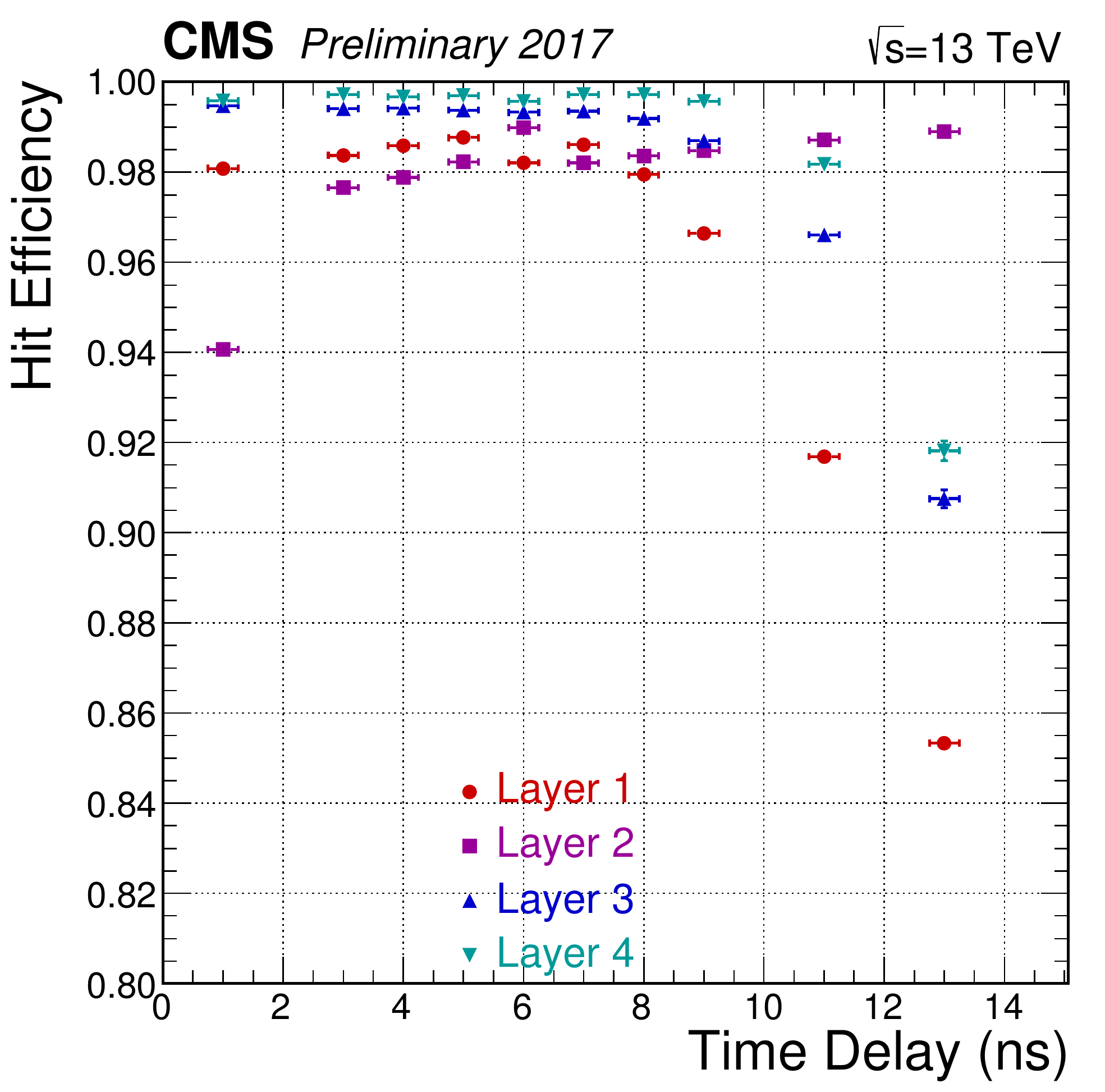}
\qquad
\includegraphics[width=.35\textwidth,origin=c]{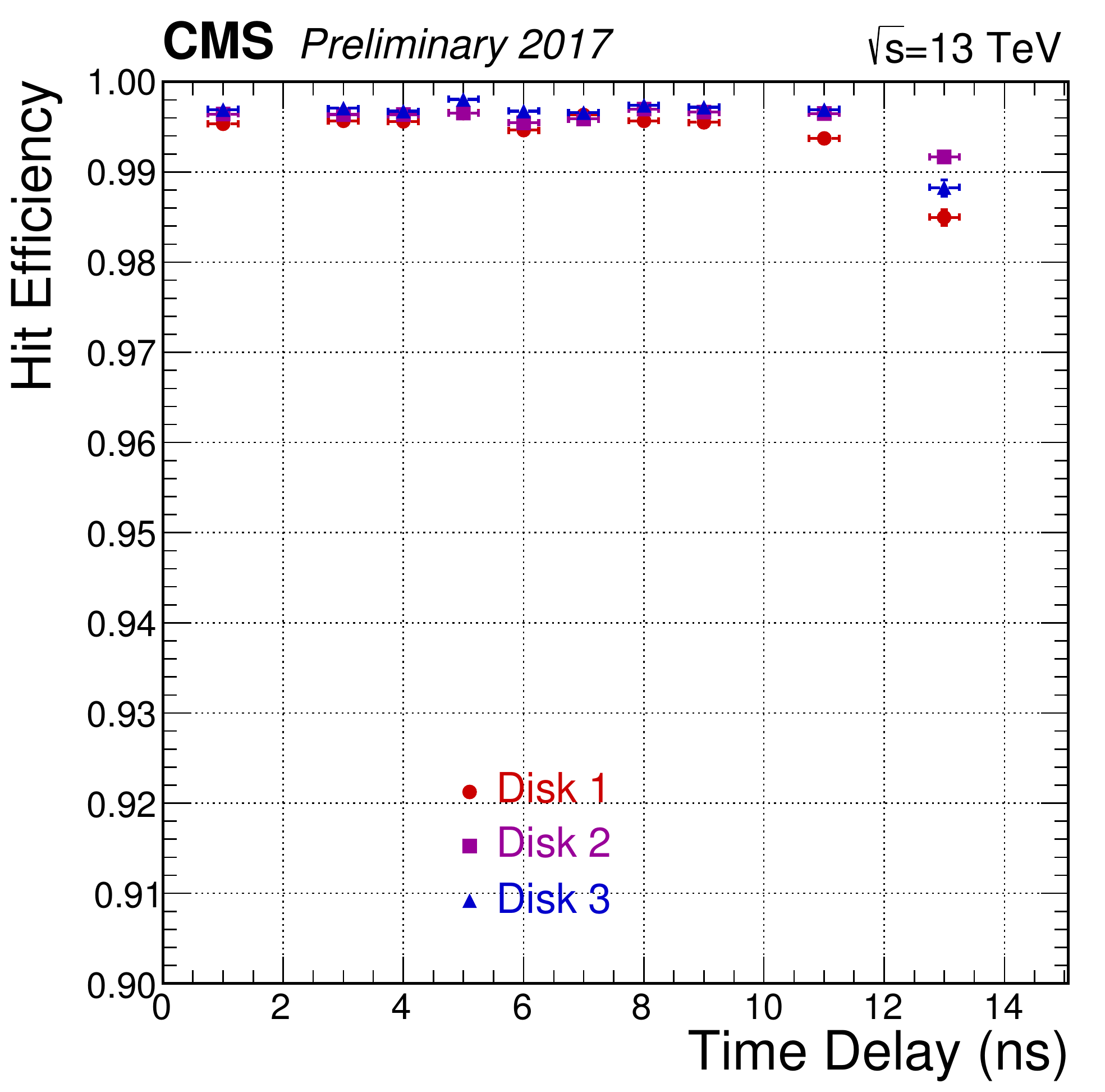}
\caption{\label{fig:DelayScanAllLayers} Left: cluster finding efficiency for hits expected along tracks in each layer of the barrel pixel detector as a function of the clock-phase delay. Right: hit efficiency in all forward disks, where disks on the two $z$-sides are averaged~\cite{CMSPixelOfflinePublicPage}. The nominal operational setting for the value of the delay is 6\,ns. }
\end{figure}

\begin{figure}[htbp]
\centering
\includegraphics[width=.35\textwidth,origin=c]{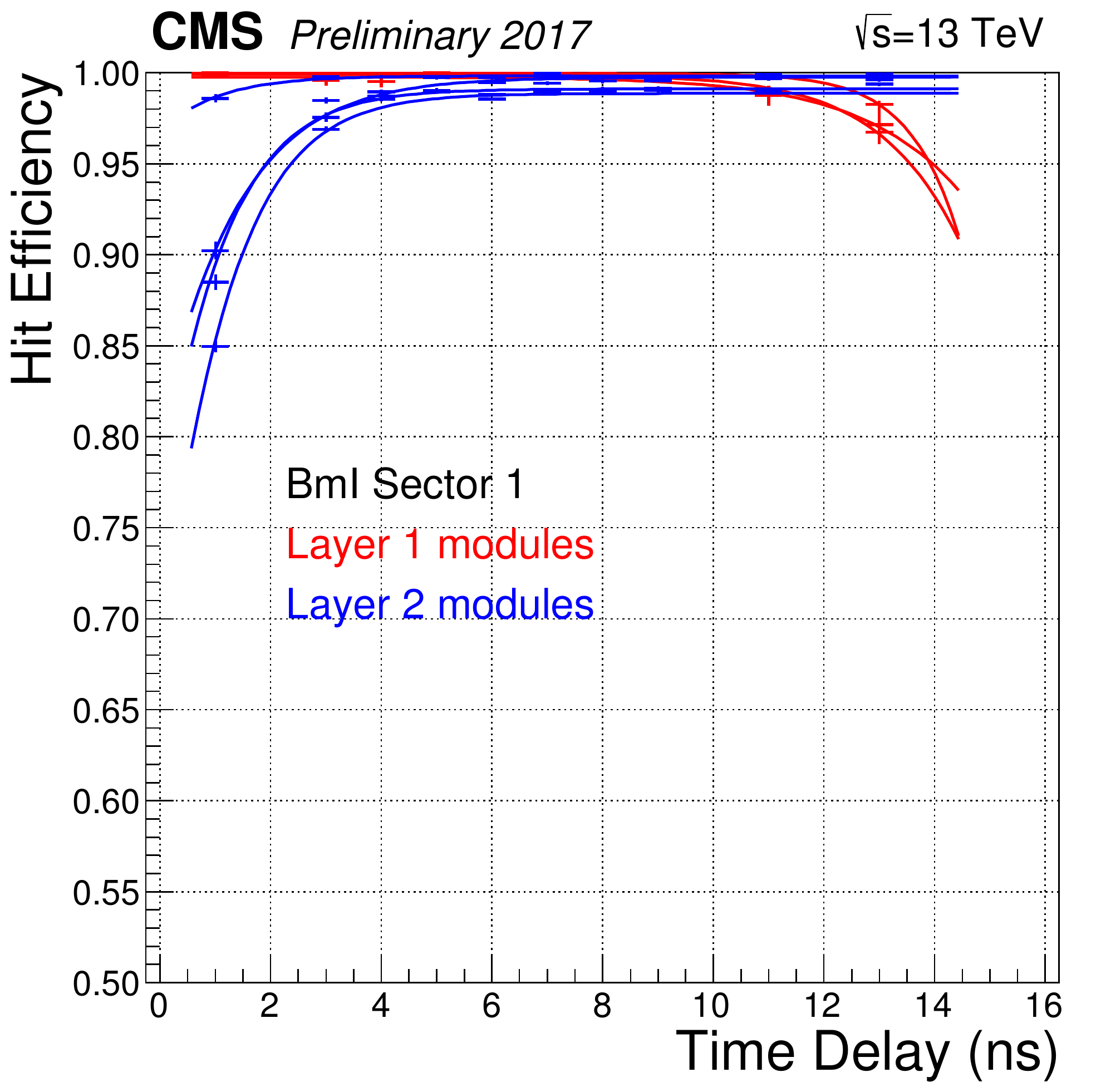}
\qquad
\includegraphics[width=.35\textwidth,origin=c]{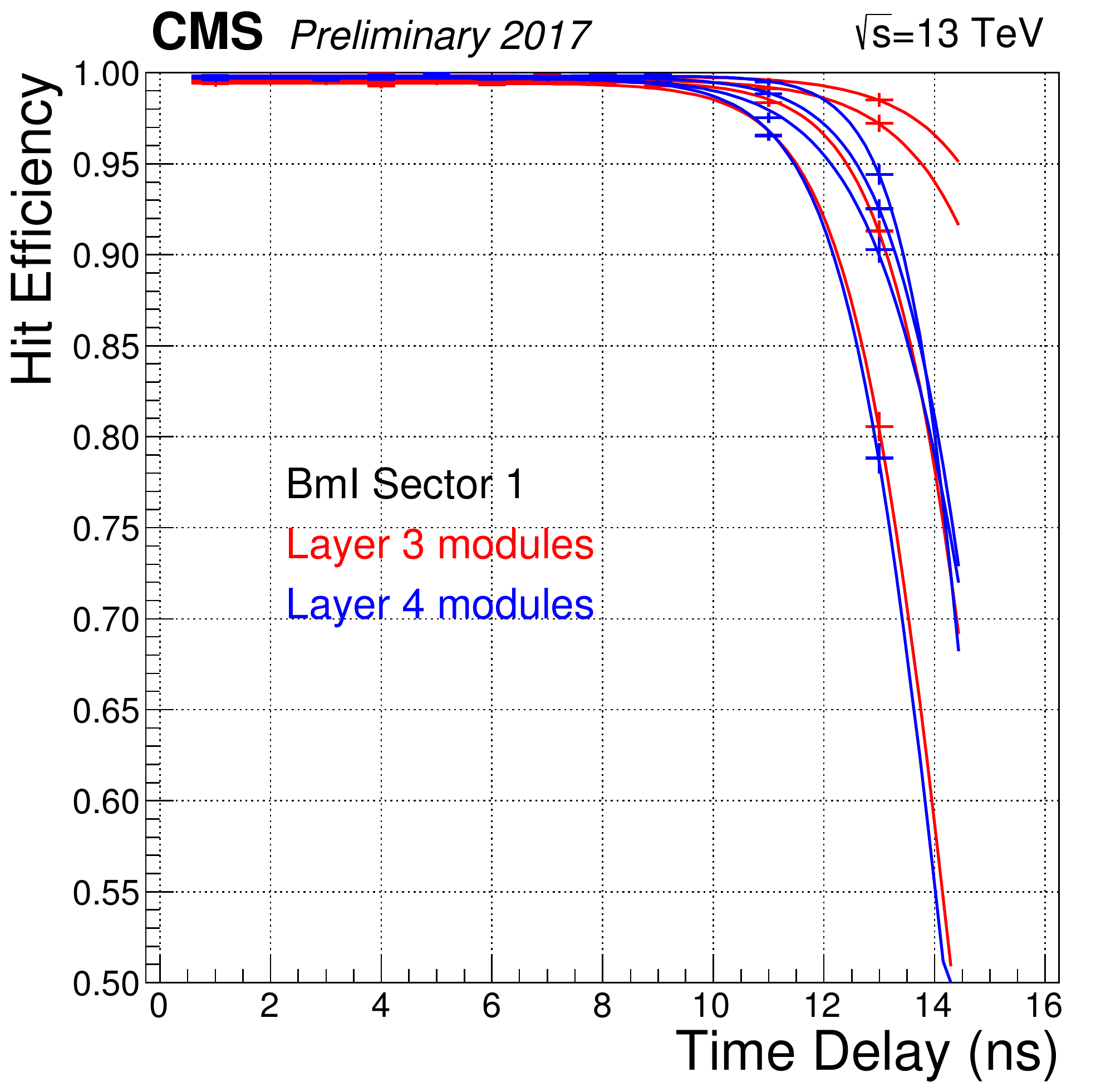}
\caption{\label{fig:DelayScanSec1} Left: hit efficiency as a function of the clock-phase delay setting for modules in the first sector of Layer~1 and 2 that belong to rings with different time of flight. Right: hit efficiency as a function of the clock-phase delay in the first sector of Layer 3 and 4~\cite{CMSPixelOfflinePublicPage}. In both cases, the clock-phase of all modules in these plots may only be programmed together. The 6 ns point was chosen for data-taking. }\end{figure}

Figure~\ref{fig:DelayScanAllLayers} shows the hit finding efficiency of all BPix layers and FPix disks as a function of the phase delay adjusted within a 15 ns window. The hit efficiency is defined as the fraction of the measurement points expected from particle trajectories propagated to the layer under test for which clusters are found within a 0.5\,mm radius area. The absolute value of the efficiency is somewhat arbitrary due to the uncertainties in tracking and the presence of temporarily malfunctioning ROCs, but the shape of the curves, with the exception of Layer~2, shows the largest delay which may be set before hits become associated to an earlier LHC bunch crossing. The goal is to delay the 25 ns window as much as possible in order to allow for the read-out of those hit pixels that are registered late but are still produced in the bunch crossing. This optimization is possible up to the accuracy of the internal synchronization of the detector. Layer~3 and 4 and the FPix disks are well adjusted. 
Figure~\ref{fig:DelayScanSec1} demonstrates an alignment consistent within approximately 3 ns for rings of modules with different time of flight values in Layer 3 and 4. The size of this spread is compatible with the per unit variations of the propagation time through the so-called LCDS chips that drive the signals on the module cables. The delay setting during data-taking is 6\,ns, a point chosen to be at least 3\,ns earlier than the falling edge of the efficiency plateau.

A more prominent difference is present among Layer~1 and Layer~2 modules. This has been identified to be due to the different phase at which the PSI46dig and the PROC600 chips associate time-stamps to hit pixels. The overlap of the feasible delay region is approximately 7\,ns, which is conveniently larger than the module-by-module spread. The delay chip of the entire Layer~1+2 clock distribution line in the DOH MB was adjusted such that the default 6\,ns delay setting optimizes for Layer~1 by keeping a good compromise for Layer~2.

\section{Calibration}

\begin{figure}[htbp]
\centering
\includegraphics[width=.35\textwidth,origin=c]{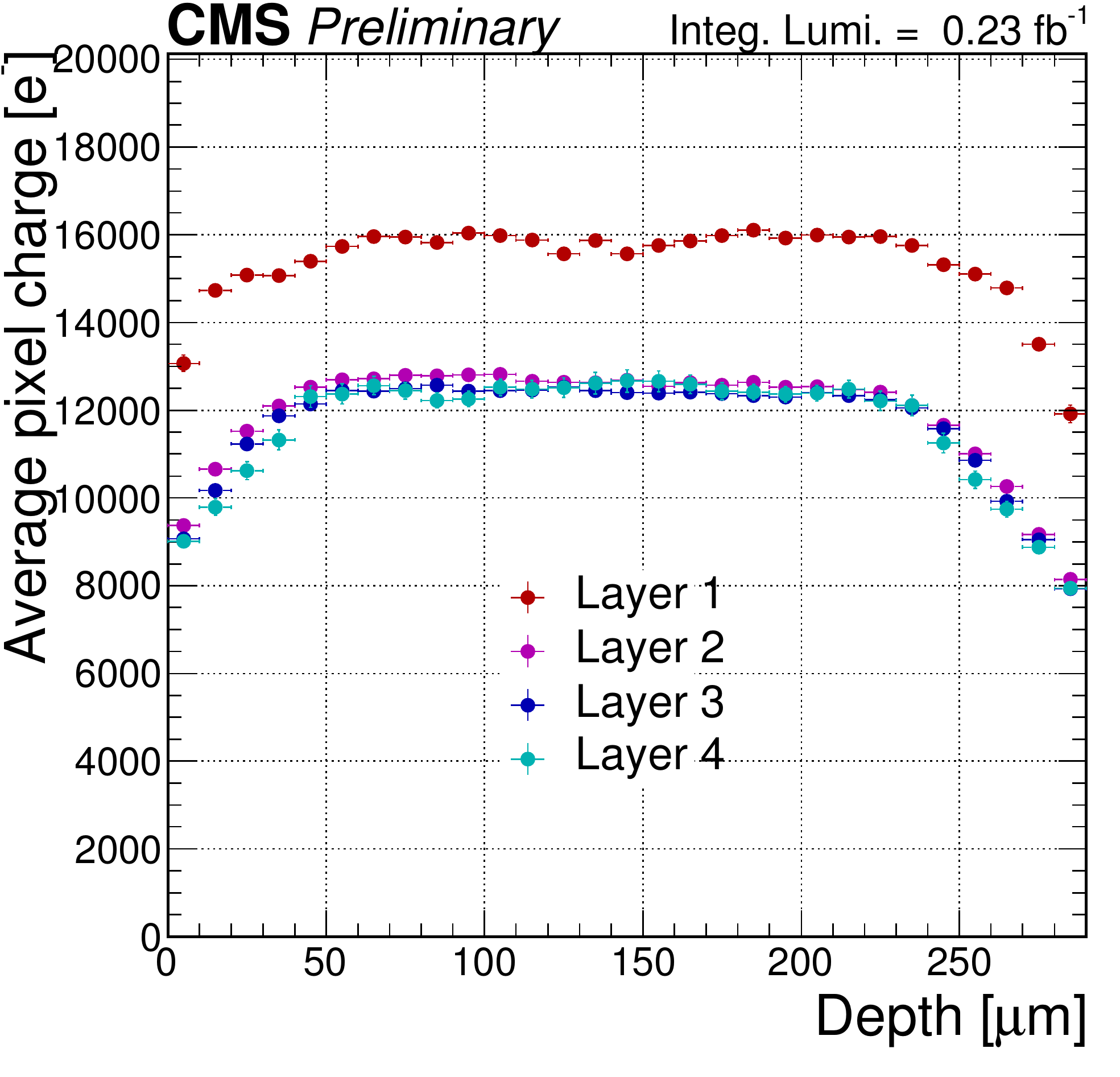}
\qquad
\includegraphics[width=.35\textwidth,origin=c]{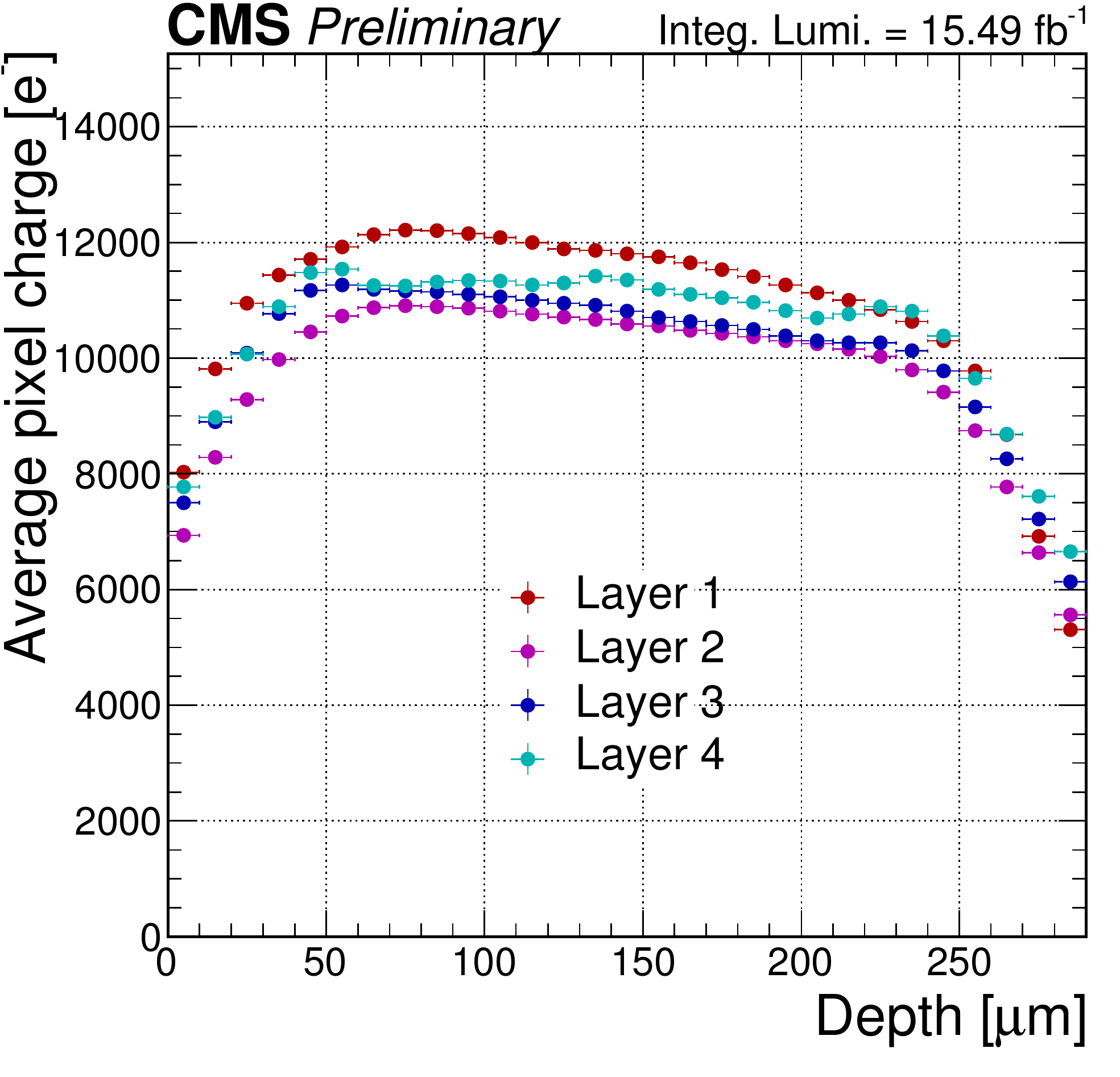}
\caption{\label{fig:ClusterChargeProfile} Average charge collected at various depths of the BPix sensors at the beginning of data-taking (left) and after collecting 15.49\,fb$^{-1}$ of data (right)~\cite{CMSPixelOfflinePublicPage}. Normalization is arbitrary: distribution of track impact angles, and temporal variations in the gain calibrations are ignored. Zero depth corresponds to the side where the pixels are read out. }
\end{figure}

The pixel charge is computed based on gain constants that are determined by charge injection. The scale of the injected charge was calibrated in X-ray measurements. After performing all the calibrations, the most probable values of the Landau-distributions for the total cluster charge are observed at around the expected value of 22\,ke$^{-}$ after normalizing all clusters to a perpendicular incidence of the incoming particles. By the end of the commissioning period, a significant amount of radiation damage and indications of charge trapping were observed that made the increase of the bias voltages necessary. Layer~1 modules are currently operated at 350\,V, Layer 2 at 250\,V, Layer 3 and 4 at 200\,V, and the FPix sensors at 300\,V. Figure~\ref{fig:ClusterChargeProfile} shows the unnormalized average charge collected from various depths of the sensors. The depth is computed from the impact angle of the track belonging to the cluster and the displacement of the pixel's center with respect to the cluster 
position. The left plot shows the shape of the distribution for a fully depleted, un-irradiated sensor, while the plot on the right shows the same after irradiation. The reduction of the charge collected from the back of the sensor was mitigated by the mentioned increase of the bias voltage.

\section{Data-taking efficiency at high hit-rate}

\begin{figure}[htbp]
\centering
\includegraphics[width=.35\textwidth,origin=c]{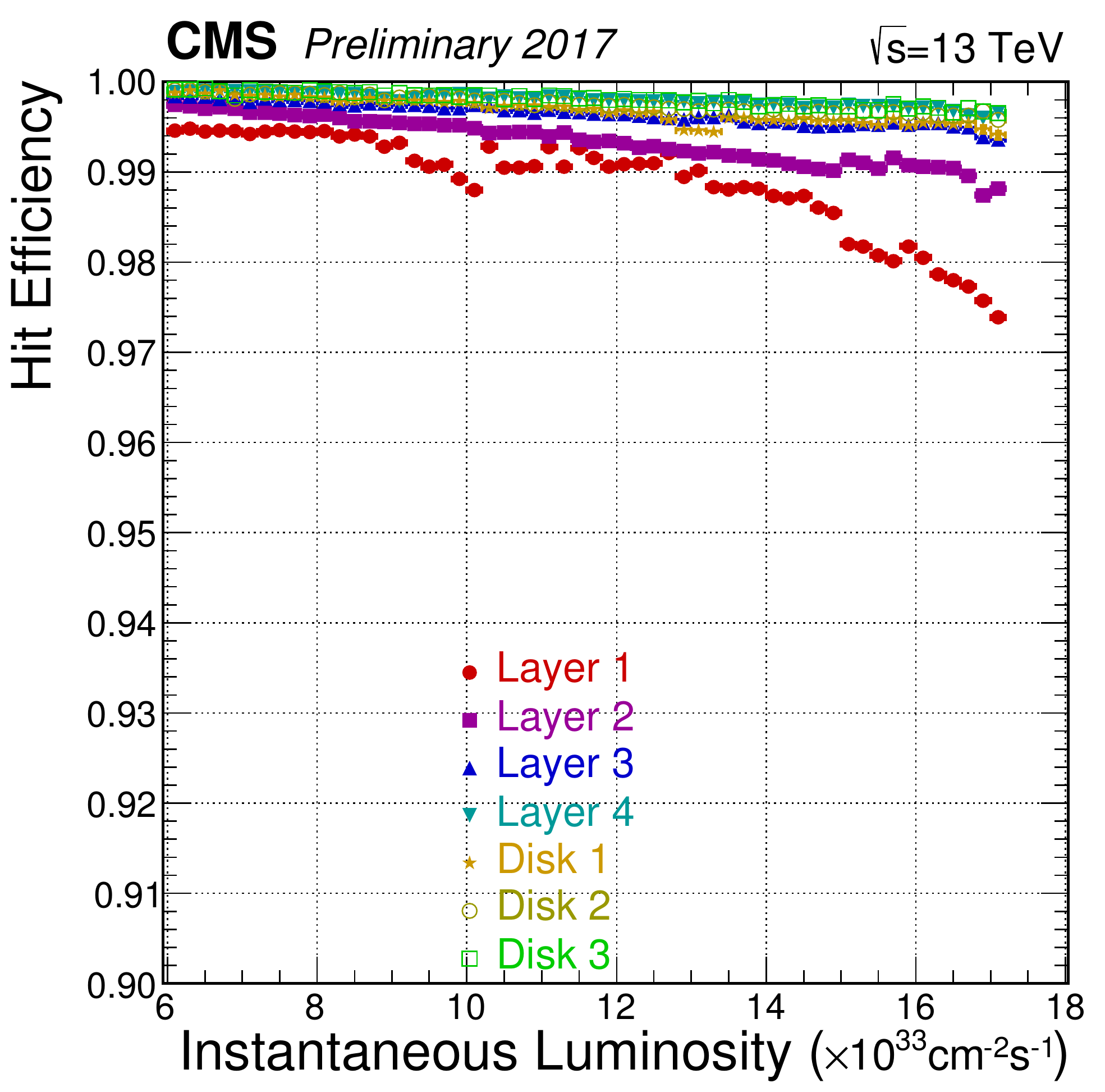}
\qquad
\includegraphics[width=.35\textwidth,origin=c]{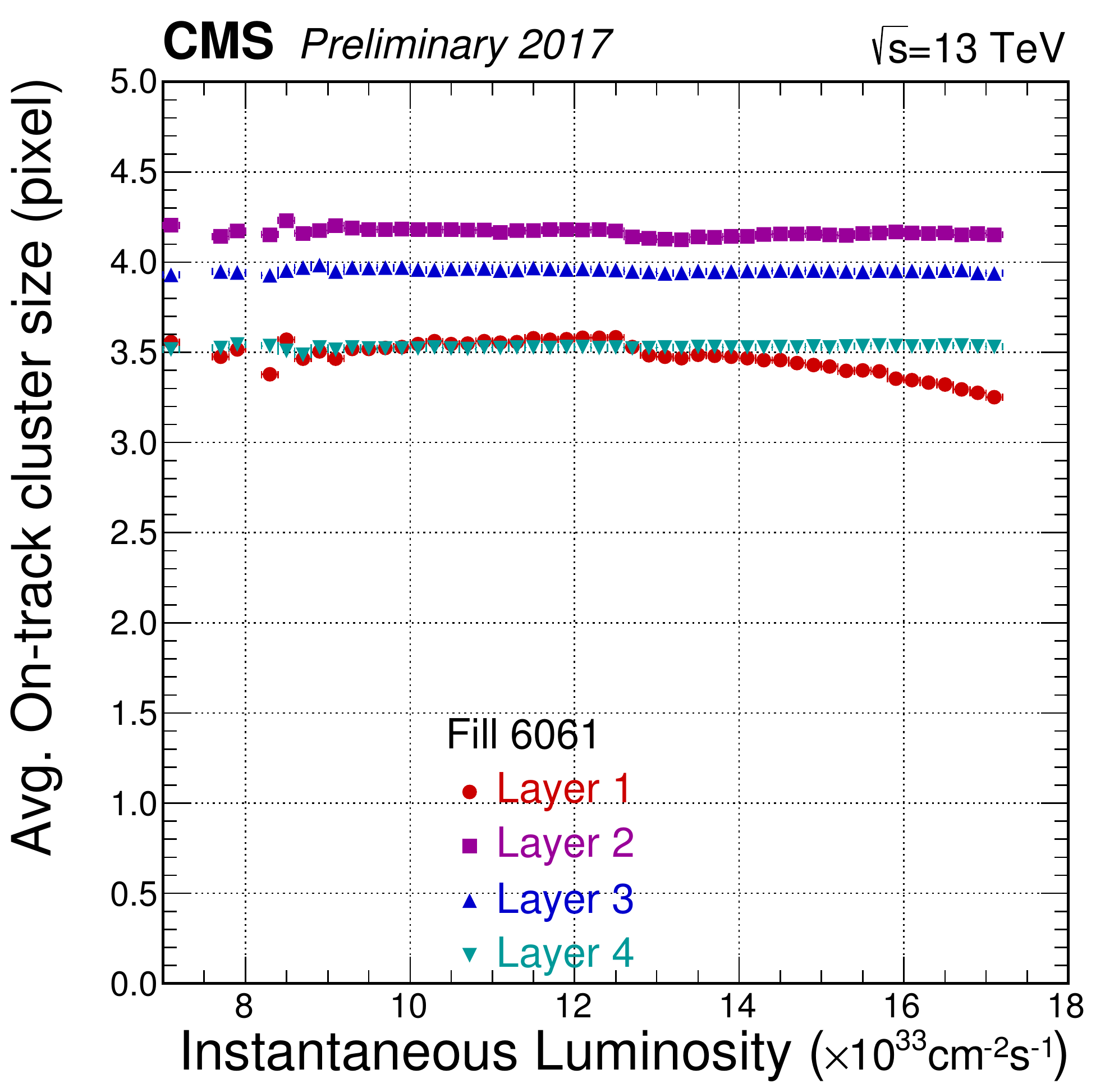}
\caption{\label{fig:EffVsInstLumi} Left: hit efficiency as a function of instantaneous luminosity for all layers and disks. A quadratic increase of the inefficiency with instantaneous luminosity is observed similarly to the 2016 detector. Right: the cluster size as a function of the instantaneous luminosity~\cite{CMSPixelOfflinePublicPage}. }
\end{figure}

As explained in Section~\ref{sec:timing}, the ROCs store pixel hits in internal buffers for the duration of the trigger latency. Time-stamp and hit buffers are placed for each double-column of the ROCs individually. In the previous version of the ROCs, the major source of inefficiency at high particle hit-rate was the overflow of these buffers. For both PSI46dig and PROC600, the size of both the time-stamp and hit buffers had been increased substantially, reducing drastically this particular contribution to the total inefficiency. While the instantaneous luminosity dependence of the hit efficiency, as shown in Figure~\ref{fig:EffVsInstLumi}, exhibits a trend similar to the old detector \cite{Phase0RadiationExperience}, the efficiency loss of 5\% observed in 2016 at 1.4$\times$10$^{34}$\,cm$^{-2}$s$^{-1}$ has reduced to about 1\%. The upgrade of the detector was absolutely necessary in order to maintain the acquisition of high quality data.

Buffer overflow leads to the loss of all hits in a double column. In the central region of the detector, where the cluster size is mostly determined by charge sharing along the direction of the double columns, this often results in the loss of the entire cluster. At higher pseudo-rapidities, where the particles arrive to the sensors at more shallow impact angles and make long clusters across multiple double columns, the net effect is cluster splitting. In both cases, measurement resolution is lost. The instantaneous luminosity dependence of the total cluster size is shown in Figure~\ref{fig:EffVsInstLumi}. 

\section{Resolution}

The advantage of the relatively thick sensor design is the possibility to profit from the Lorentz-force induced charge sharing in order to increase the measurement resolution. Since the magnetic field is parallel to the global $z$ direction of the CMS reference frame, and the electric field in the BPix sensors is approximately aligned radially in the transverse plane, the direction of the charge sharing is tangential to the surface of the sensor pointing in the r-$\phi$ direction. The size of a pixel in this direction is 100\,$\mu$m, but the measurement resolution is an order of magnitude better. In the FPix, charge sharing is induced by rotating the plane of the modules by 20$\degree$ about the radial axis. In addition, in the inner rings, blades are tilted out of the transverse plane by 12$\degree$. The complex geometry of the FPix leads to eight different configurations of the electric field with respect to the magnetic field lines.

\begin{figure}[htbp]
\centering
\includegraphics[width=.35\textwidth,origin=c]{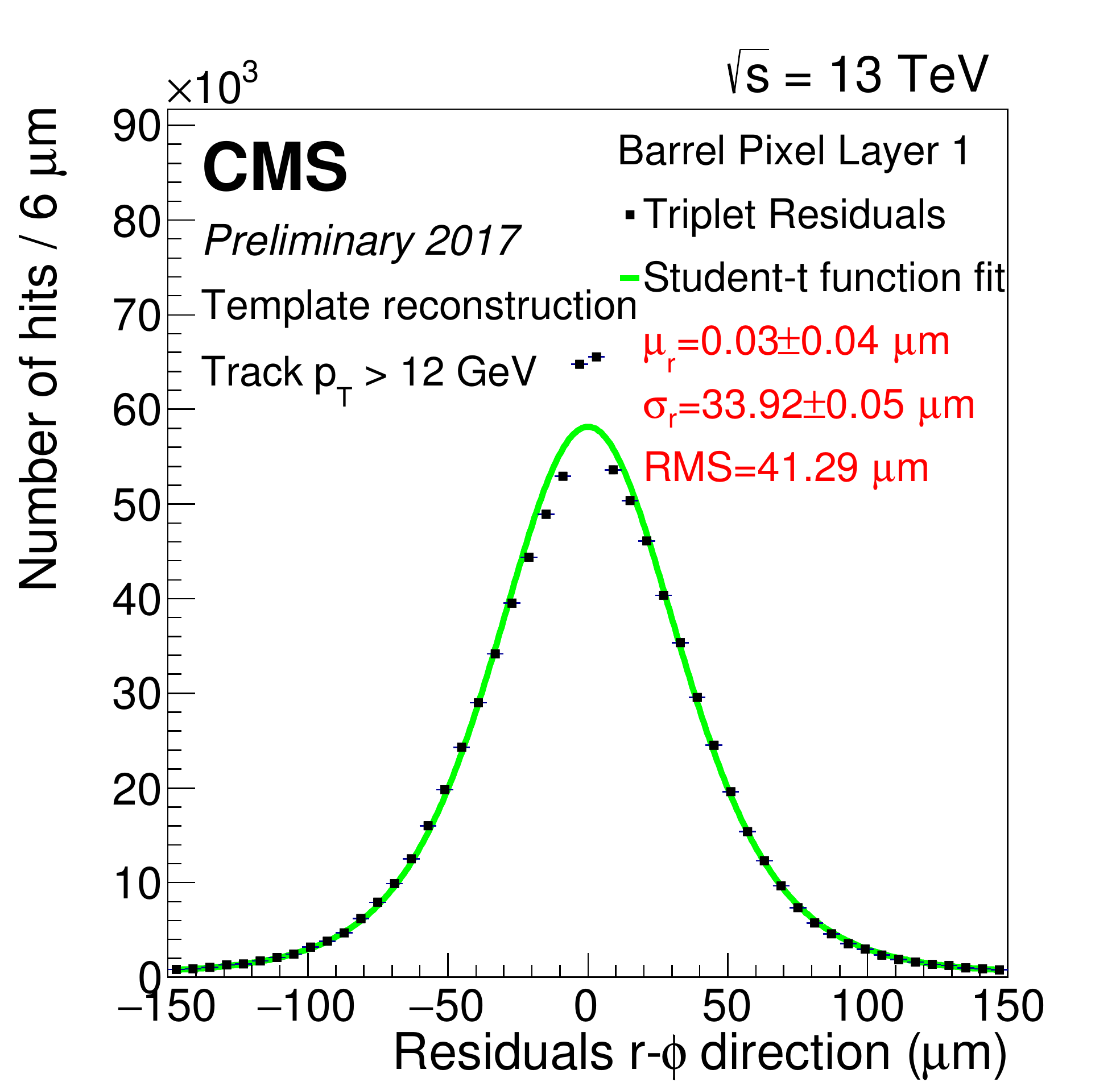}
\qquad
\includegraphics[width=.35\textwidth,origin=c]{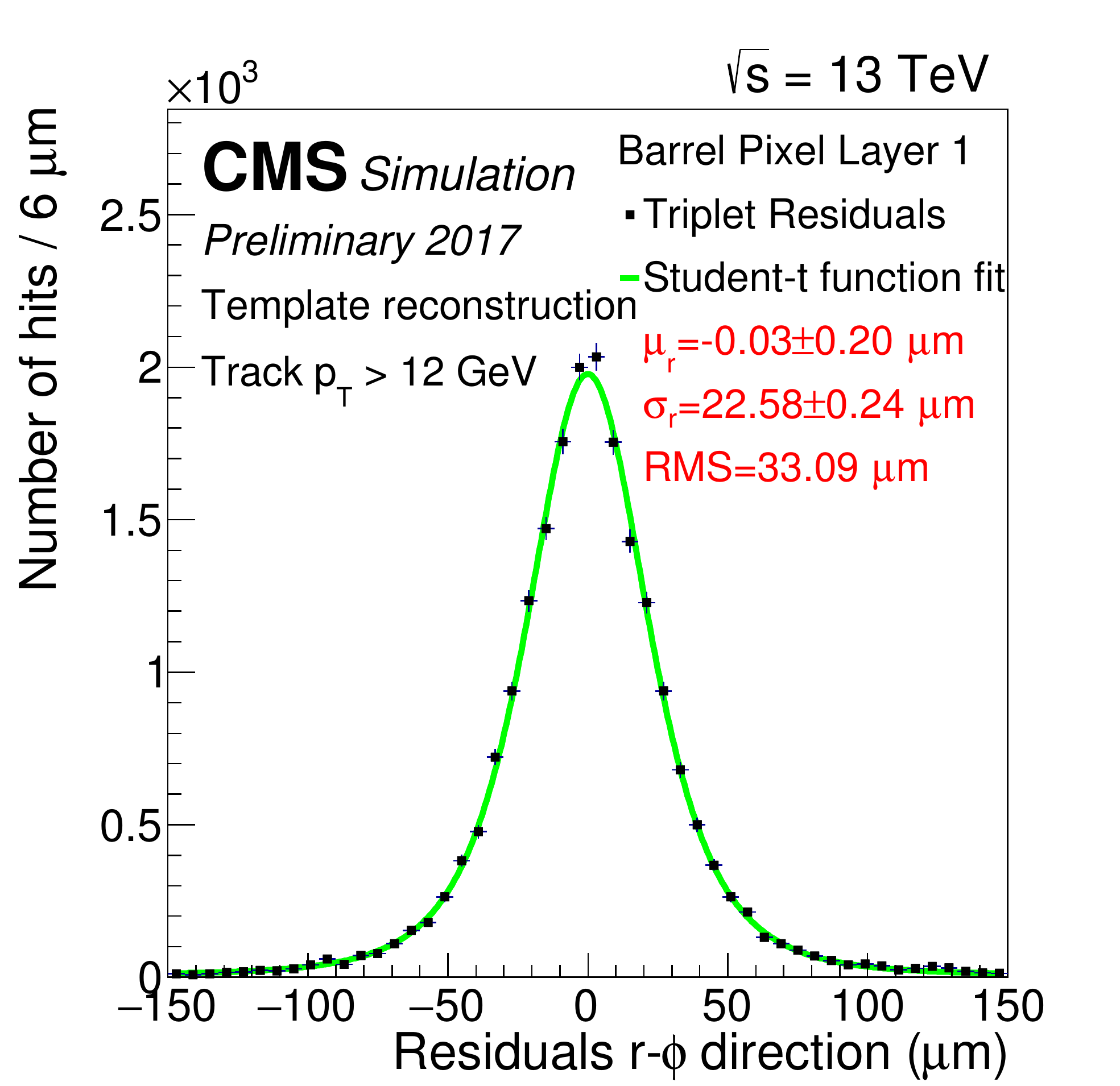}
\caption{\label{fig:ResidualLayer1} Hit triplet residual measurement in Layer~1 from data (left) and simulation (right) along the r-$\phi$ direction, in the transverse plane~\cite{CMSPixelOfflinePublicPage}. }
\end{figure}

\begin{figure}[htbp]
\centering
\includegraphics[width=.35\textwidth,origin=c]{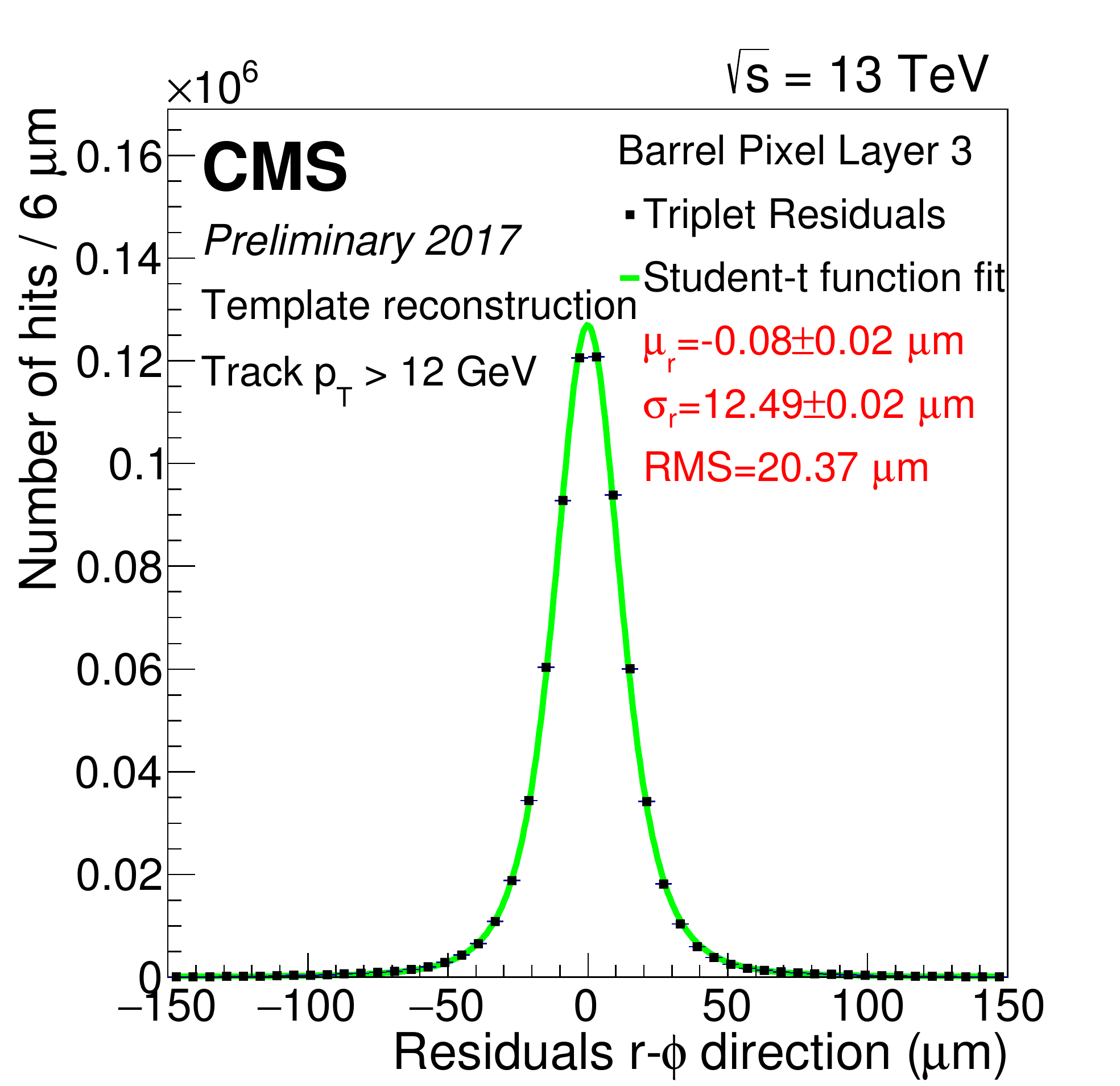}
\qquad
\includegraphics[width=.35\textwidth,origin=c]{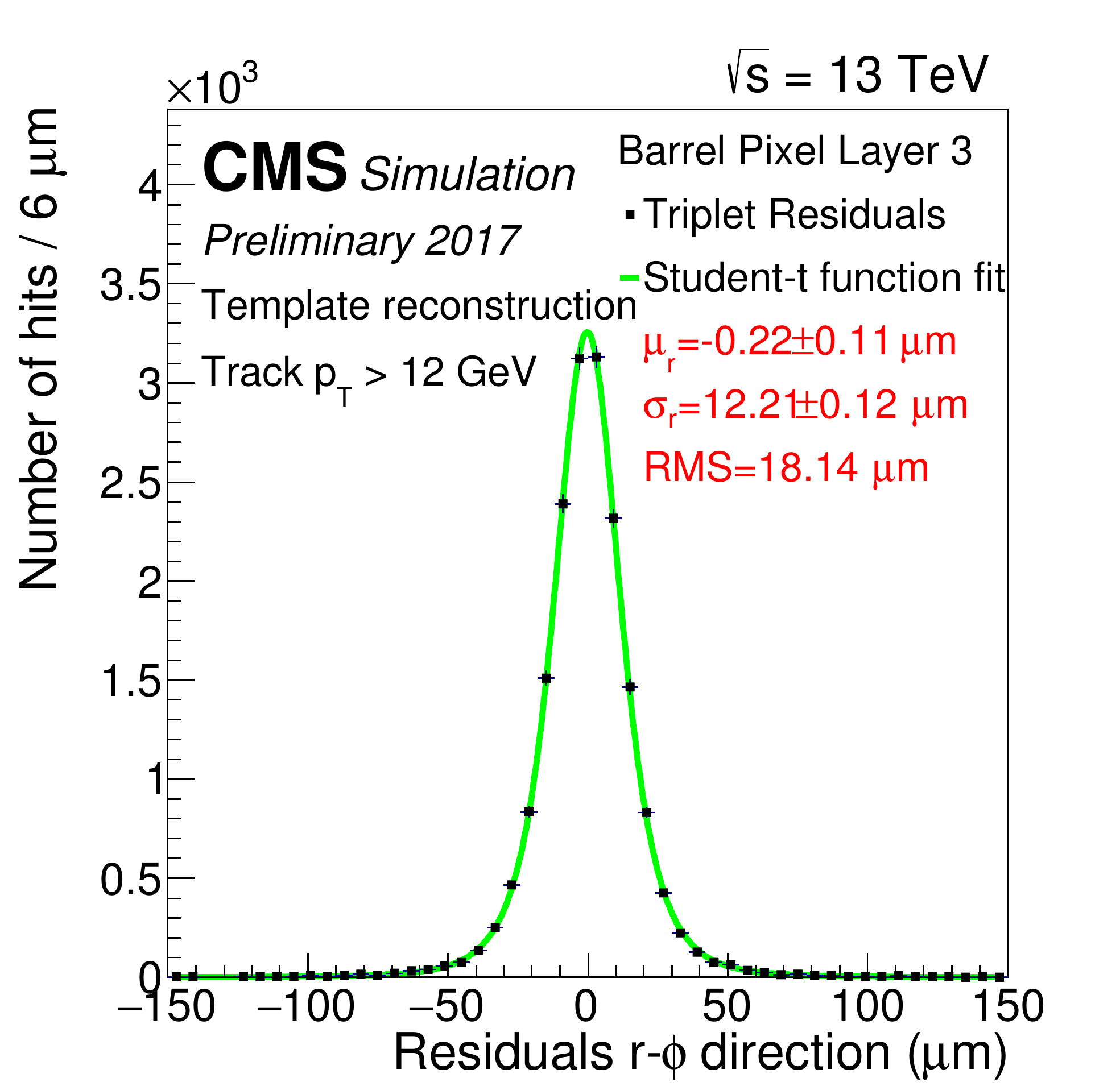}
\caption{\label{fig:ResidualLayer3} Hit triplet residual measurement in Layer 3 from data (left) and simulation (right) along the r-$\phi$ direction, in the transverse plane~\cite{CMSPixelOfflinePublicPage}. }
\end{figure}

The resolution is qualitatively determined by the measurement of hit residuals along well reconstructed tracks (Figures~\ref{fig:ResidualLayer1} and~\ref{fig:ResidualLayer3}.) Tracks with at least three hits in three different pixel layers or disks are selected in the first step of the measurement. The residuals for one hit in the triplet is computed as the difference between the extrapolated trajectory state and the position of the cluster attached to the hit. The trajectory is propagated analytically from the hit doublet using the momentum of the original track. The position of the cluster is determined by the template fit method~\cite{Pixelav}.

Figures~\ref{fig:ResidualLayer1} and ~\ref{fig:ResidualLayer3} show the residuals for Layer~1 and 3 as examples. Such measurements have been performed for all layers in BPix and all the different kind of module positions in FPix. In the figures, the plots on the left show results from collision data. The residual not only does reflect the position error of the cluster on the layer under test but also the error of the other two cluster positions in the triplet propagated to the measurement by tracking. At this point of the commissioning, there is no reliable estimate to compute such propagated errors, but the results may be compared to the same measurement performed in Monte Carlo simulations. Monte Carlo results are shown on the right. At the end of the commissioning process, the simulation reproduced the residuals in all layers and disks (as demonstrated for Layer 3) with the exception for Layer~1. In these measurements, the simulation assumes no efficiency loss due to high particle rate, no radiation 
damage, and reasonably low, 2\,ke$^{-}$ thresholds. Comparison of the reconstructed hits with simulated hits in Monte Carlo implies that all layers and disks reached their ultimate intrinsic resolution. The Layer~1 resolution is reduced by a higher than expected threshold, the reduced charge collection efficiency, double column inefficiency, and not yet perfectly calibrated cluster position measurement.

\section{Summary and outlook}

The upgraded Phase-1 detector has been successfully installed and commissioned in 2017. The preliminary results of the detector performance at the end of the commissioning process are presented in this paper. The primary goal for the upgrade, to ensure the acquisition of high-quality data at the increased instantaneous luminosity of the LHC, is met. The new detector is taking data with sensor efficiency better than 99\% up to an instantaneous luminosity twice as high as when this threshold was reached last year; although, the new detector is showing somewhat lower efficiency at high hit-rate then initially expected. The data rate limitations in the DAQ system have been extended approximately to the target of the LHC up to 2023. Since the sensor design has not changed, the same hit resolution is observed as that of the old detector. However, an improved track reconstruction has been made possible by the extra layers in the detector at no cost on the material budget or on read-out limitations. A few challenges 
have also been encountered. The Layer~1 modules are operated at a threshold higher than expected due to cross-talk. There is a clock-phase difference between the PSI46dig and PROC600 chips that made the time alignment of the detector more difficult, but with no penalty on the final performance.

The performance of Layer~1 is expected to improve with more elaborate calibration procedures towards the end of 2017, which will also be beneficial in the reprocessing of this year's data for physics analyses.

\end{document}